\documentclass{article}

\usepackage{arxiv}
\usepackage{hyperref}        % hyperlinks
\usepackage{siunitx}		  % comprehensive (SI) units 
\usepackage{booktabs}       % professional-quality tables
\usepackage{amsfonts}       % blackboard math fonts
\usepackage{amsmath}       % blackboard math symbols
\usepackage{multirow}       % multirow / multicolumn table
\usepackage{graphicx}       % graphic tools
\usepackage{pifont} 

\title{SenAOReFoc: A Closed-Loop Sensorbased Adaptive Optics and Remote Focusing Control Software}

\author{
\textbf{Jiahe Cui} \\
Department of Engineering Science \\ 
University of Oxford \\ 
Parks Road, Oxford, OX1 3PJ, UK \\
jiahe.cui@eng.ox.ac.uk 
\and
\textbf{Karen M. Hampson} \\
Department of Engineering Science \\ 
University of Oxford \\ 
Parks Road, Oxford, OX1 3PJ, UK \\
karen.hampson@eng.ox.ac.uk
\AND
\textbf{Matthew Wincott} \\
Department of Engineering Science \\ 
University of Oxford \\ 
Parks Road, Oxford, OX1 3PJ, UK \\
matthew.wincott@eng.ox.ac.uk
\and
\textbf{Martin J. Booth}\\
Department of Engineering Science \\ 
University of Oxford \\ 
Parks Road, Oxford, OX1 3PJ, UK \\
martin.booth@eng.ox.ac.uk
}

\begin{document}
\maketitle

\begin{abstract}
SenAOReFoc is a closed-loop sensorbased adaptive optics (AO) and remote focusing control software that works with a deformable mirror (DM) and a Shack-Hartmann wavefront sensor (SHWS). It is programmed in Python and is open-source on Github \url{https://github.com/jiahecui/SenAOReFoc}. A detailed user guide can be found in the Github repository. Here, we give a brief summary of basic software functionalities, a statement of need, and some examples for the usage of this software.
\end{abstract}

\keywords{adaptive optics \and closed-loop \and sensorbased \and remote focusing \and control software}

\section{Summary}
SenAOReFoc is a closed-loop sensorbased adaptive optics (AO) and remote focusing control software that works with a deformable mirror (DM) and a Shack-Hartmann wavefront sensor (SHWS). It provides a  user-friendly graphic user interface (GUI) with modular widget arrangements and clear labelling to help the user navigate through different software functionalities. Interactive messages are also displayed from the GUI for user guidance.

\begin{figure}[h!tb] 
	\centering
	\includegraphics[width=1.0\linewidth]{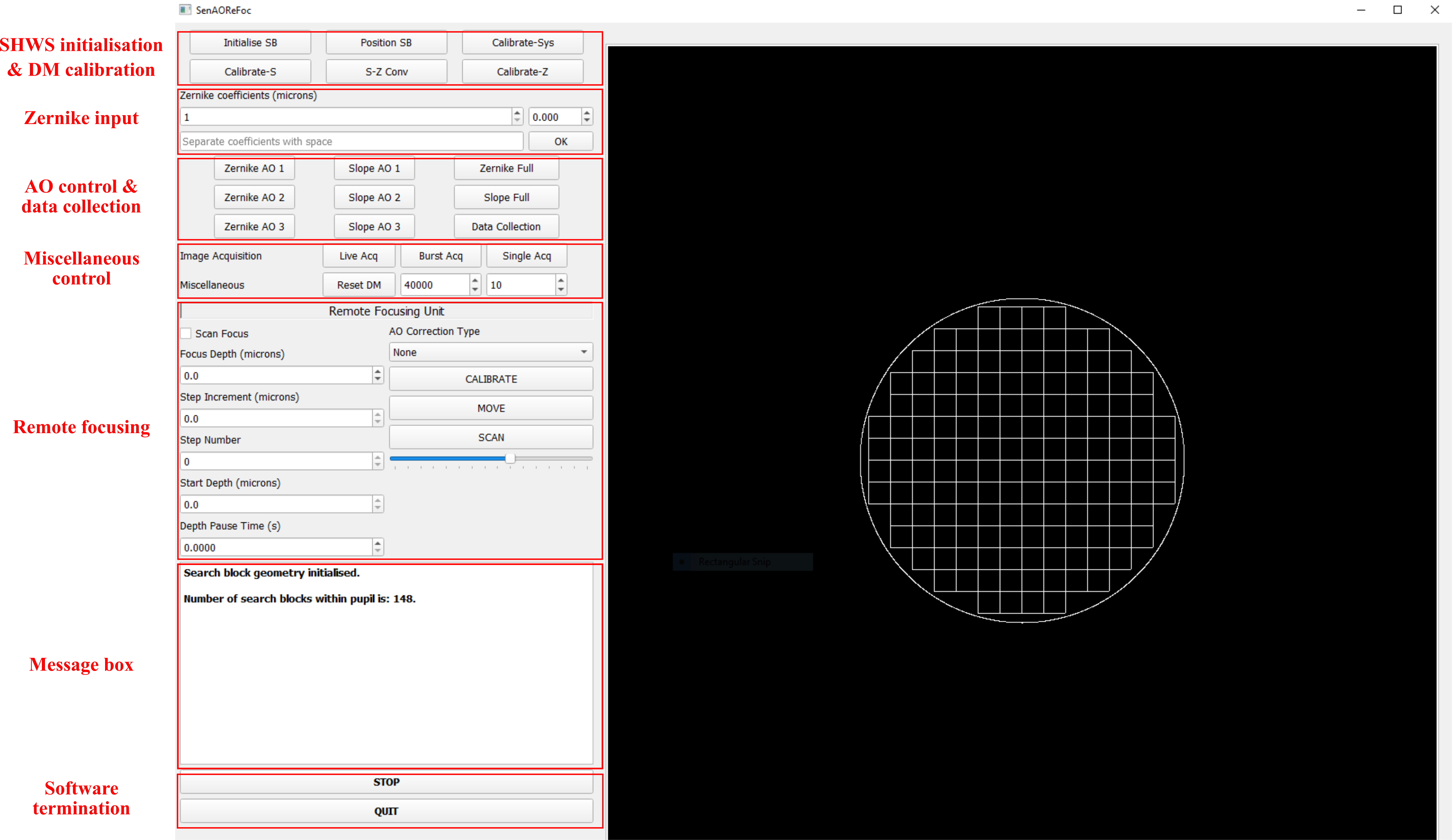}
	\caption{Graphic user interface (GUI) of SenAOReFoc with modular units labelled in red boxes.}
	\label{fig:gui}
\end{figure}

SenAOReFoc consists of 5 main units, the SHWS initialisation and DM calibration unit, the Zernike aberration input unit, the AO control and data collection unit, the miscellaneous control unit, and the remote focusing unit, as shown in Figure~\ref{fig:gui}. The software can be ran in either `debug mode' to perform functionality tests without connected hardware (DM and SHWS), or `standard mode' on a well-aligned optical sectioning microscope (confocal, multiphoton, etc.). User controllable system parameters can be freely accessed and modified in a separate configuration file that is loaded upon software initialisation, and parameters that require continuous user input can be modified from the GUI. Parameters calculated when running the software, as well as important result data, are grouped and saved in a separate HDF5 file that can be read with HDFView software. Automated AO performance characterisations can be performed in `standard mode' to assess the correction ability of the optical system. If the adopted DM is designed with a large stroke, i.e., is capable of large deformations, both the closed-loop AO correction and remote focusing functionalities can be exploited. On the other hand, if the DM exhibits insufficient stroke for remote focusing, by ignoring the remote focusing unit, closed-loop AO correction functionalities will still be fully functional without additional modifications to the software. 

Closed-loop AO correction can be performed using both the zonal method, which updates DM control voltages in terms of the raw slope values; and the modal method, which updates DM control voltages in terms of orthogonal Zernike polynomials. There are four sub-modes tagged to each of the two methods: 1) standard closed-loop AO correction; 2) closed-loop AO correction with consideration of obscured search blocks; 3) closed-loop AO correction with partial correction excluding defocus; and 4) closed-loop AO correction with both consideration of obscured search blocks and partial correction excluding defocus.

Remote focusing can be performed by scanning the focus axially with a pre-determined axial range, step increment and step number, or by manually adjusting a toggle bar on the GUI for random access remote focusing. The former also incorporates options of whether or not to perform closed-loop AO correction at each remote focusing depth.

%%%%%%
%%%%%%

\section{Statement of need}
The performance of optical microscopes degrades significantly in the presence of optical aberrations that develop due to misalignments in the optical system and inhomogeneities in refractive index of the imaging sample. This results in compromised image resolution and contrast. Adaptive optics (AO) is a powerful technique that corrects for optical aberrations using reconfigurable adaptive devices, such as a deformable mirror (DM) or spatial light modulator (SLM), to restore the image quality~\cite{Booth2014LightScience&Applications}. This is most crucial for high-quality imaging, such as super-resolution or deep imaging.

Remote focusing is an axial scanning technique recently introduced to avoid slow movements of the sample stage and objective lens during real-time applications~\cite{Ji2016NatureNeuroscience}. By using a DM for remote focusing, calibration can be performed through closed-loop AO correction to simultaneously correct for system aberrations introduced by beam divergence at different focusing depths~\cite{Zurauskas2017BiomedicalOpticsExpress}

Sensorbased AO, which uses a dedicated sensor to measure the wavefront~\cite{Platt2001JournalofRefractiveSurgery}, was first introduced in the field of astronomy~\cite{Beckers1993}. Since then, it has also been widely adopted in vision science~\cite{Porter2006,Godara2010Optometryandvisionscience} and microscopy~\cite{Booth2007PhilosophicalTransactions,Ji2017NatureMethods}. A range of software serving for different astronomical purposes have been introduced over time, including those for simulation of AO systems~\cite{Carbillet2005MonthlyNoticesoftheRoyalAstronomicalSociety,Conan2014}, control with atmospheric tomography~\cite{Ahmadia2005}, and fast control using GPU hardware~\cite{Guyon2018}. In the microscopy community, software is also openly available for AO modelling and analysis~\cite{Townson2019OpticsExpress}, as well as general AO control~\cite{Hall2020OpticsExpress}. However, there is not yet an existing open-source software that serves the purpose of performing both AO and remote focusing within the single package, despite that the techniques have been widely adopted. As a result, time and effort has to be spent reimplementing existing techniques for different hardware systems. 

SenAOReFoc aims to fill this gap and to make closed-loop sensorbased AO~\cite{Fernandez2001OpticsLetters} and remote focusing more easily accessible to the microscopy community. It has been designed for open development and easy integration into existing adaptive optical microscopes with a simple and user-friendly architecture. The functionality of the software has also been tested on different operating systems (Windows/macOS/Linux) for sake of generality. However, we note that SenAOReFoc is a control software, and the performance of the closed-loop AO correction in practice is inevitably dependent on the state of the optical system.
Finally, SenAOReFoc takes care of some reported issues in the field, such as the thermal effects of electromagnetic DMs~\cite{Bitenc2017OpticsExpress}, and obscured search blocks in the case of severely distorted pupils~\cite{Dong2018OpticsExpress,Ye2015OpticsExpress,Cui2020Zenodo}.

\section{Example usage}
Four examples are given for the usage of this software, the first three for automated AO performance characterisations of the reflectance confocal microscope described in~\cite{Cui2021Biophotonics,Cui2021OpticsExpress}; and the last for remote focusing in frozen mouse skull as reported in~\cite{Cui2021Biophotonics}.

%%%%%%
\begin{figure}[h!tb] 
	\centering
	\includegraphics[width=0.6\linewidth]{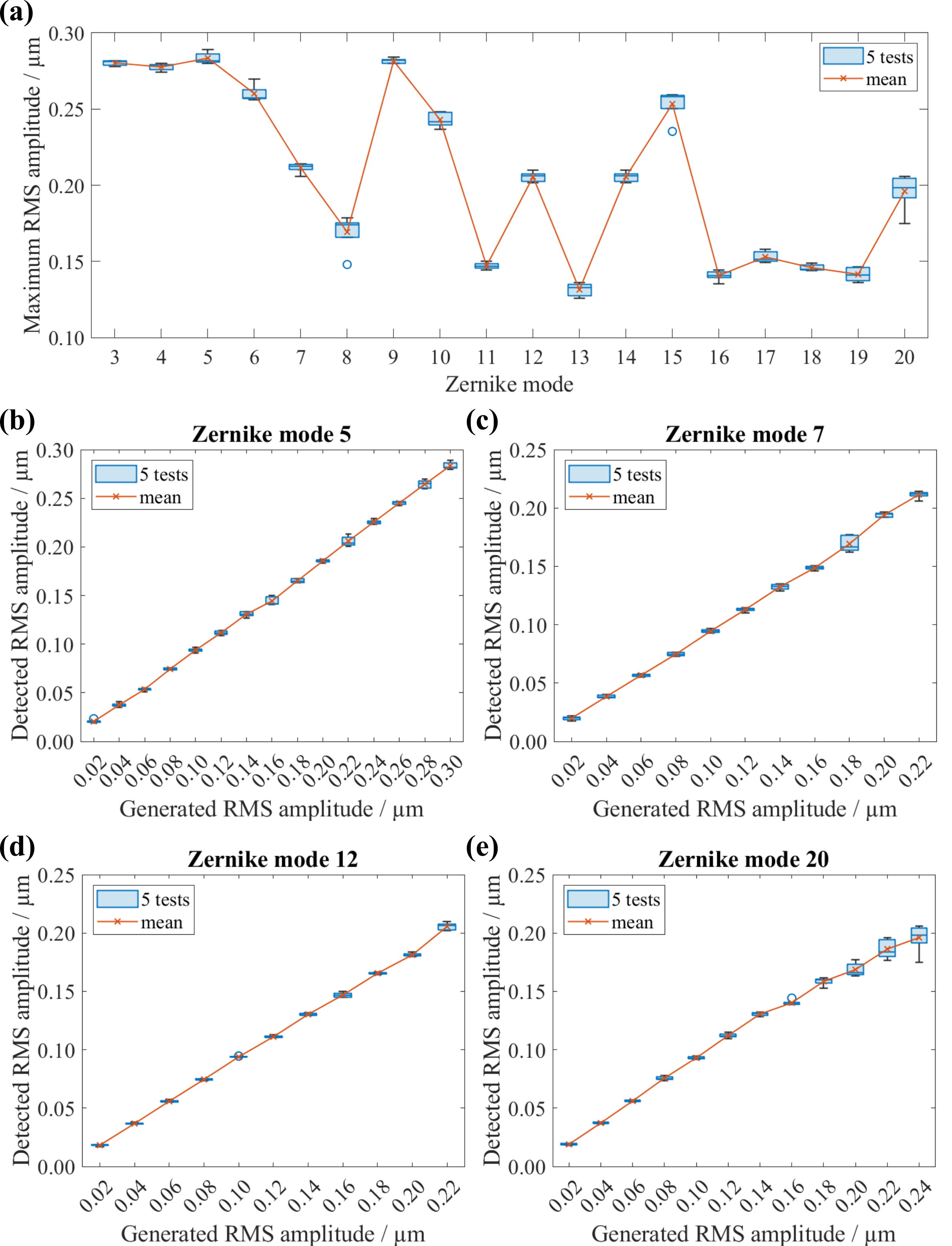}
	\caption{Characterisation results of system AO correction performance. (a) Dynamic range of the SHWS for Zernike modes 3-20 (excl. tip/tilt). (b)-(e) Generated and detected RMS amplitudes of odd/even Zernike modes (b) 5, (c) 7, (d) 12, and (e) 20, in increments of 0.02 micrometers. 5 tests were performed for each measurement.}
	\label{fig:example1}
\end{figure}
%%%%%%

\textbf{Example 1}: The dynamic range of the SHWS for the first \textbf{N} Zernike modes can be characterised by generating and correcting for them in closed-loop using mode 0/1 of [Data Collection]. Figure~\ref{fig:example1}(a) provides an example when characterising Zernike modes 3-20 (excl. tip/tilt). Figure~\ref{fig:example1}(b)-(e) provide examples of generated and detected RMS amplitudes of selected odd/even Zernike modes in increments of 0.02 micrometers. Parameters in \verb|config.yaml| under \verb|AO| and \verb|data_collect| were set as follows.

\begin{verbatim}
AO:
control_coeff_num: 20  # Number of zernike modes to control during AO correction

data_collect:
data_collect_mode: 1  # Mode flag for Data Collection button, detailed function 
                        descriptions in app.py
loop_max_gen: 15  # Maximum number of loops during closed-loop generation of zernike 
                    modes in mode 0/1/2/3
incre_num: 15  # Number of zernike mode amplitudes to generate in mode 0/1
incre_amp: 0.02  # Increment amplitude between zernike modes in mode 0/1
run_num: 5  # Number of times to run mode 0/1/2/3
\end{verbatim}

\begin{figure}[h!tb] 
	\centering
	\includegraphics[width=0.5\linewidth]{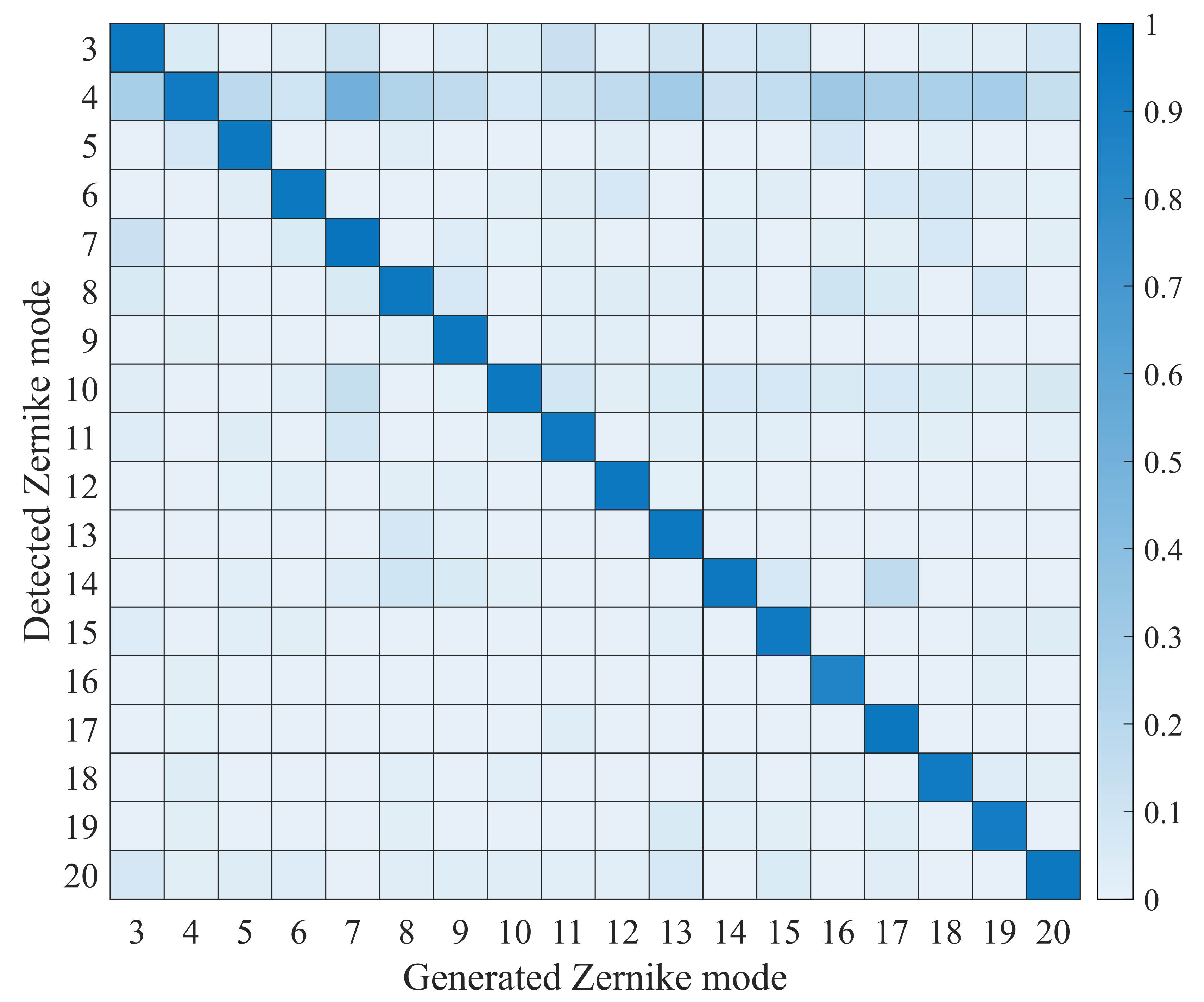}
	\caption{Heatmap of correlation coefficients between detected and generated mode values for 0.1 micrometers of Zernike modes 3-20 (excl. tip/tilt).}
	\label{fig:example2}
\end{figure}

\textbf{Example 2}: The degree of Zernike mode coupling upon detection at the SHWS can be characterised by individually generating the same amount of each mode on the DM and constructing a heatmap of the detected Zernike coefficients using mode 0/1 of [Data Collection]. Figure~\ref{fig:example2} provides an example heatmap of correlation coefficients between detected and generated mode values for 0.1 micrometers of Zernike modes 3-20 (excl. tip/tilt). Parameters in \verb|config.yaml| under \verb|data_collect| were set as follows.

\begin{verbatim}
data_collect:
data_collect_mode: 1  # Mode flag for Data Collection button, detailed function 
                        descriptions in app.py
loop_max_gen: 15  # Maximum number of loops during closed-loop generation of zernike
                    modes in mode 0/1/2/3
incre_num: 1  # Number of zernike mode amplitudes to generate in mode 0/1
incre_amp: 0.1  # Increment amplitude between zernike modes in mode 0/1
run_num: 5  # Number of times to run mode 0/1/2/3
\end{verbatim}

%%%%%%
%%%%%%

\begin{figure}[h!tb] 
	\centering
	\includegraphics[width=1.0\linewidth]{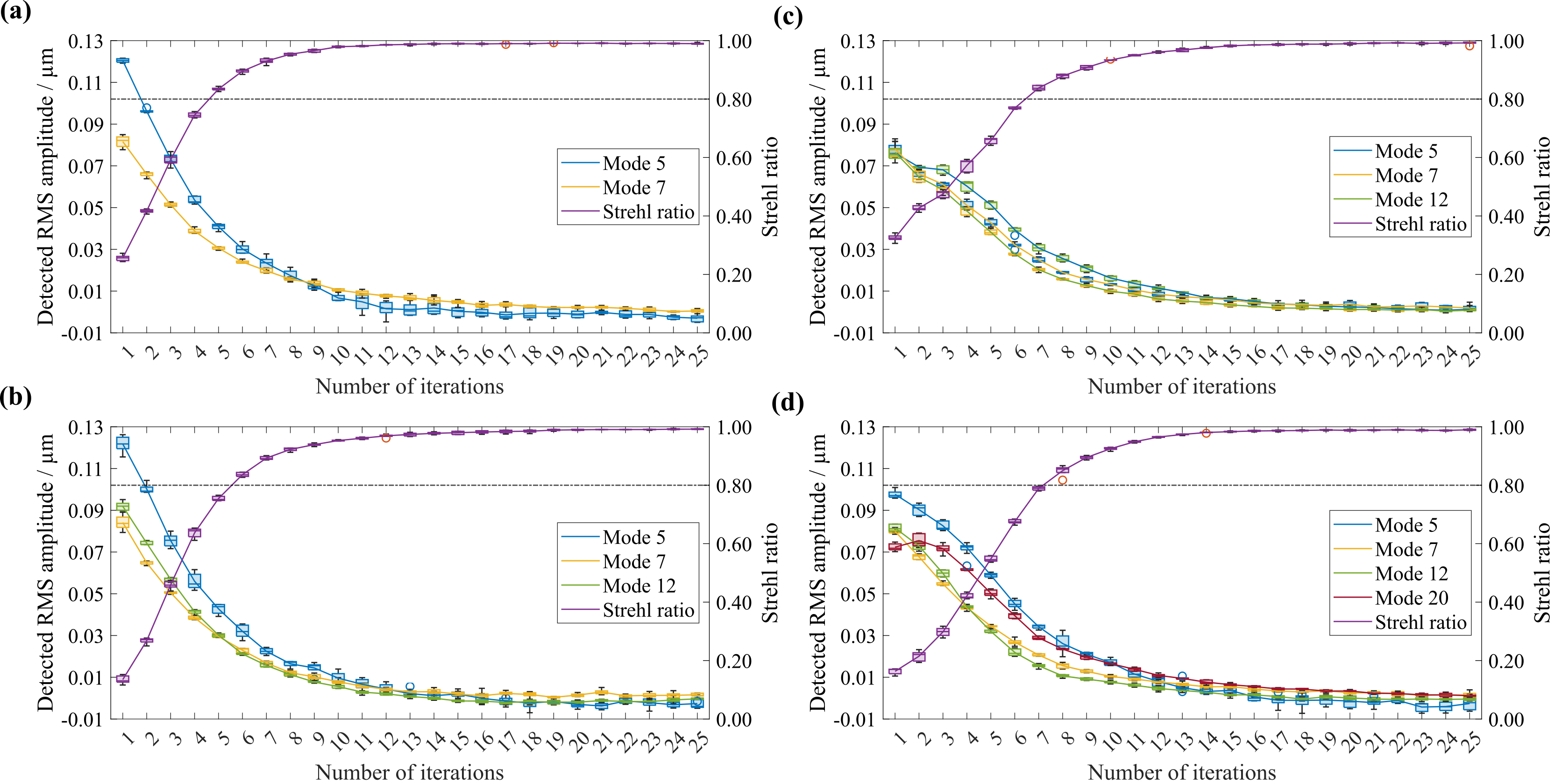}
	\caption{Detected amplitudes of generated odd and even Zernike mode combinations and Strehl ratio calculated using first 69 Zernike modes (excl. tip/tilt) after each closed-loop iteration. 5 tests were performed for each measurement.}
	\label{fig:example3}
\end{figure}

\textbf{Example 3}: To ensure the system can correct for multiple Zernike modes with good stability and minimal mode coupling, different combinations of odd and even Zernike modes can be generated and corrected for in closed-loop using mode 2/3 of [Data Collection]. Figure~\ref{fig:example3} provides an example of the detect amplitude and Strehl ratio after each closed-loop iteration for some odd and even Zernike mode combinations. Parameters in \verb|config.yaml| under \verb|data_collect| were set as follows.

\begin{verbatim}
data_collect:
data_collect_mode: 3  # Mode flag for Data Collection button, detailed function 
                        descriptions in app.py
loop_max_gen: 15  # Maximum number of loops during closed-loop generation of zernike 
                    modes in mode 0/1/2/3
run_num: 5  # Number of times to run mode 0/1/2/3
\end{verbatim}

And [Zernike array edit box] was set to [0 0 0 0 \textbf{0.1} 0 \textbf{0.1} 0 0 0 0 \textbf{0.1} 0 0 0 0 0 0 0 \textbf{0.1}].

%%%%%%
%%%%%%

%%%%%%
\begin{figure}[h!tb] 
	\centering
	\includegraphics[width=1.0\linewidth]{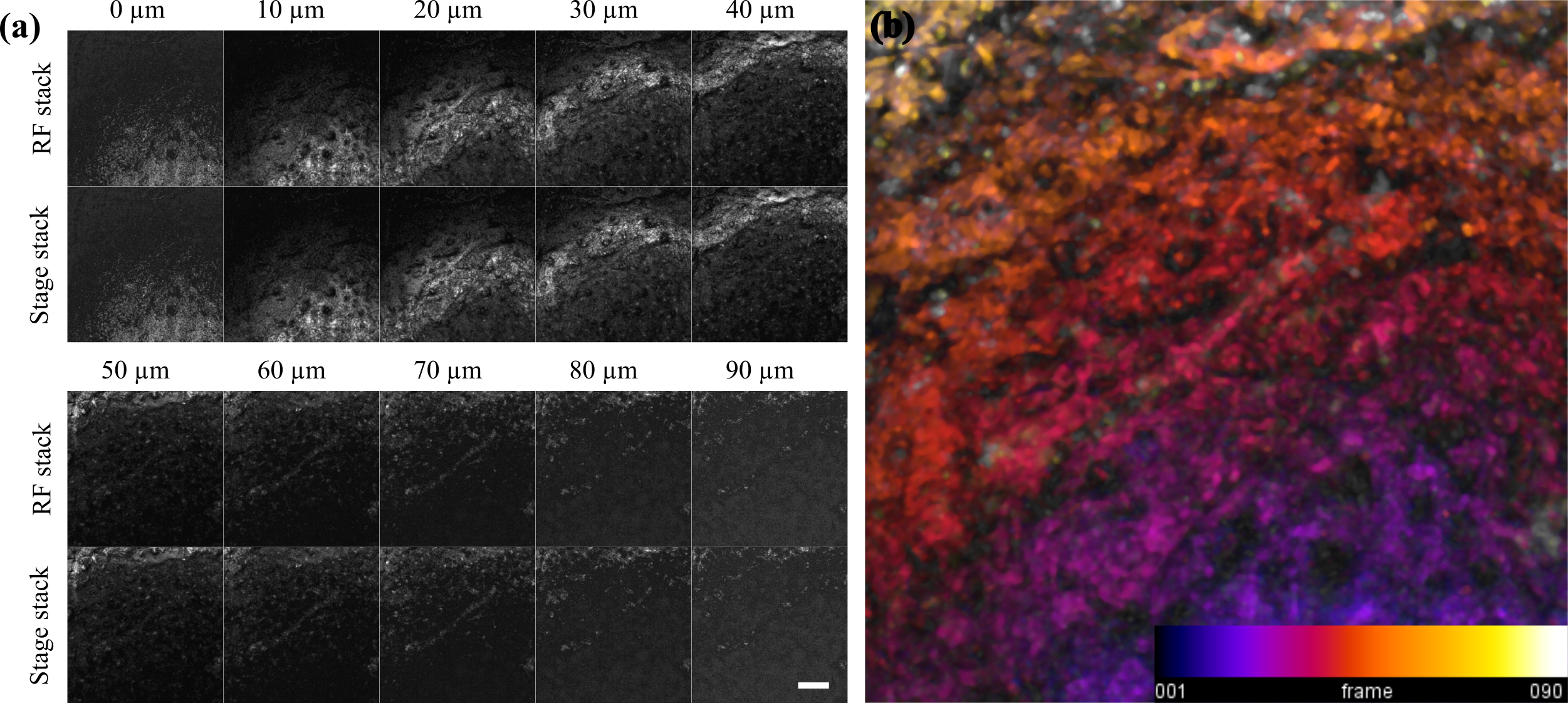}
	\caption{Remote focusing results within frozen mouse skull. (a) Stack images acquired with 10 micrometer axial steps over a 90 micrometer depth range. Top rows: results obtained by scanning sequentially through multiple depths using pre-calibrated DM control voltages. Bottom rows: comparison results obtained by axially translating the sample stage only. (b) Colour-coded maximum intensity projection of 90 frames with 1 micrometer intervals. Scale bar: 100 micrometers.}
	\label{fig:example4}
\end{figure}
%%%%%%

\textbf{Example 4}: Voltages that control DM actuators to deform the membrane for fine axial remote focusing was calibrated according to the procedure explained in the GitHub repository \url{https://github.com/jiahecui/SenAOReFoc}, and reported in~\cite{Cui2021Biophotonics}and~\cite{Cui2021OpticsExpress}. [Calibrate] was pressed for both directions of the optical axis and a piece of white card was displaced by 10 micrometers each time before pressing \verb|y| on the keyboard to proceed with closed-loop AO correction. Interpolation of the DM control voltages at each calibration step was then performed to obtain those for 0.1 micrometer increments. The software's remote focusing capability was demonstrated in frozen mouse skull in~\cite{Cui2021Biophotonics} to show no noticeable difference in
resolution and size of field of view as compared to standard translation of the sample stage. Results are also shown in Figure~\ref{fig:example4}.

\section*{Acknowledgements}
We thank Chao He for testing the software functionality using different operating systems. This work was supported by European Research Council projects 695140 and 812998.

\bibliographystyle{ieeetr}  
\bibliography{references}  

\end{document}